\documentclass{agujournal}
\draftfalse

\journalname{JGR-Planets}
\usepackage{array}

\begin{document}

%
%


\title{Examining MAVEN NGIMS Neutral Data Response to Solar Wind Drivers}

%
%




\authors{H. N. Williamson\affil{1}, M. K. Elrod\affil{2}, S. M. Curry\affil{3}, R. E. Johnson\affil{1}}


\affiliation{1}{University of Virginia}
\affiliation{2}{NASA Goddard Spaceflight Center}
\affiliation{3}{Space Sciences Laboratory, UC Berkeley}




\correspondingauthor{Hayley Williamson}{hnw9ew@virginia.edu}




\begin{keypoints}
\item We examine MAVEN NGIMS neutral data for evidence of sputtering due to ion precipitation
\item The data is converted to Mars-Solar-Electric coordinates so location indicates solar wind direction
\item Density changes in MSE coordinates are suggestive of an effect but need more data to be conclusive
\end{keypoints}

%
%


\begin{abstract}
The Martian upper atmosphere is known to vary diurnally and seasonally due to changing amounts of solar radiation. However, in the upper thermosphere and exosphere, the neutrals are also subject to ion precipitation. This can increase the temperature in the region of precipitation, resulting in density changes that might be seen in \textit{in situ} data \citep{Fang2013}. Therefore, we examine neutral density data from the MAVEN Neutral Gas and Ion Mass Spectrometer (NGIMS) in Mars-Solar-Electric (MSE) coordinates, where location is determined by the direction of the solar wind convective electric field, resulting in a hemispherical asymmetry in the ion precipitation. By examining densities in MSE coordinates \citep{Hara2017} we are able to look for a detectable effect in the region where ion precipitation is more likely. Using the NGIMS neutral data and Key Parameters \textit{in situ} solar wind data from February 2015 to August 2017 we look for asymmetries by constructing average density maps in Mars-Solar-Orbital (MSO) and MSE coordinates near the exobase. The NGIMS densities for O, Ar, and $\mathrm{CO_2}$ from 180-220 km altitude for each orbit are averaged and then binned by location in MSO coordinates and transformed to MSE coordinates. The resulting MSE map exhibits a small density increase in the southern hemisphere, where one would expect to see enhanced precipitation. Although suggestive, the change is not statistically significant, so that the effect of ion precipitation, thought to be an important driver in the evolution of Mars' atmosphere remains elusive.

\end{abstract}

%
%

%


%
%
%
%

\section{Introduction} \label{intro}
The Mars Atmosphere and Volatile EvolutioN (MAVEN) mission seeks to both understand the structure of Mars' current atmosphere and ascertain how much of this atmosphere has been lost to space over time, known as atmospheric escape \citep{Jakosky2015}. The Sun, and subsequently the solar wind, a stream of charged particles that flows off of the Sun, can affect the atmosphere and drive escape in several ways. For example, the interplanetary magnetic field (IMF) carried in the solar wind will pick up ions in the ionosphere, which can then be lost by flowing down the magnetospheric tail \citep{Curry2015}. However, while many of the ions picked up by the IMF are accelerated away from the planet, some reenter the atmosphere in a process known as pick-up ion precipitation \citep{Johnson1998}. The precipitating ions can then collide with and transfer energy to the neutral atmosphere. Non-thermal collisions can result in sputtering, a splashing out effect on the neutrals which has been suggested to be responsible for much of early neutral atmospheric escape \citep{Leblanc2002}. However, it is now estimated to be a very small part of the present escape rate and as such is difficult to directly detect in MAVEN data \citep{Leblanc2015}. Since the energy transfer from precipitating ion and neutral collisions can raise the temperature of the atmosphere, and hence the scale height, leading to an increase in density, here we use the changes in the average density in the exobase region as a proxy for the pick-up ion heating. Precipitation is an ongoing process even during quiet solar conditions \citep{Leblanc2015,Hara2017}, so in this paper we look at average densities at a given altitude range to see if there is at present an effect from precipitation. \\

The neutral densities in this paper were obtained by the Neutral Gas and Ion Mass Spectrometer (NGIMS) on MAVEN, a quadrupole mass spectrometer that measures \textit{in situ} densities every orbit. NGIMS provides a unique opportunity to study upper atmospheric composition, as it has now measured densities between the nominal altitudes of approximately 150-350 km for over a full Martian year. This provides a large dataset for analysis of the neutral densities in the altitude regime where the atmosphere transitions from collisional to ballistic and neutral escape becomes more likely. Since ion precipitation is correlated with the direction of the solar wind convective electric field, it is more likely when the field is directed towards the planet \citep{Brain2015,Fang2013,Hara2017}. Thus examining neutral densities in a frame of reference dependent on the direction of said electric field, the Mars-Solar-Electric Field coordinates (MSE), can help in trying to understand how and if ions driven by the solar wind can contribute to neutral escape. We also use Mars-Solar-Orbital (MSO) coordinates, which are useful for both atmospheric data and data taken farther from the planet, such as in the upstream solar wind, unlike geodetic coordinates, as a control comparison.\\

In MSE coordinates, the 'southern' latitudes indicate the electric field is pointed towards the planet, while 'northern' latitudes indicate the electric field is pointed away from the planet. Therefore precipitation predominately occurs in the southern MSE, or -E hemisphere, whereas ions flow out in the northern MSE latitudes or +E hemisphere. Transforming the neutral data to this coordinate system allows us to examine the effect on the neutrals of the potential presence of ion precipitation. We do this by comparing measured densities where precipitation is occurring with densities where precipitating ions are probably not providing additional energy to the neutral atmosphere.\\

 We calculate average densities for oxygen, argon, and carbon dioxide between 180 and 220 km, then map these averages in MSO coordinates. We then use the upstream solar wind proton flow velocity and IMF vectors to calculate the solar wind convective electric field direction and use this to find MSE coordinates. As such, we will presume that changes in density in this altitude range between the northern and southern hemispheres are most likely due to the transfer of energy from precipitating ions in the southern -E hemisphere. The absence of a clear effect would suggest that ion precipitation is not affecting the neutrals. We also look at the data for different solar longitudes, to compare with global circulation models of the neutral atmosphere in order to ascertain if there is a seasonal effect. While we do not see any clear difference between seasonal densities in the data and previous model results (e.g. \cite{Bougher2015}), we do see a suggestion that average neutral densities are slightly higher in areas where precipitation likely occurs. Although the evidence is not statistically significant, it is suggestive and could become clearer with additional data.

\section{Methods} \label{methods}
This study uses the publicly available NGIMS level 2, version 7, revision 3 data from February 2015 to August 2017, slightly more than a full Martian year, for a total of 3828 orbits. NGIMS is a quadrupole mass spectrometer that measures \textit{in situ} ion and neutral counts for a range of 2-150 amu with 1 amu resolution every orbit. Both ion and neutrals counts are measured by NGIMS in channels for mass-to-charge ratio in a 2.6 s cadence \citep{Benna2017}. The counts from each mass channel are converted to abundances in particles per cubic centimeter vs. altitude, time, latitude, and longitude of the measurement in the instruments level 2 data files.\\

MSO coordinates refer to a Mars-fixed solar-pointing coordinate system, with the X vector pointing towards the Sun, the Y vector anti-parallel to the direction of the orbit, and the Z vector completing the orthogonal system \citep{vignes2000}. In the MSE coordinate system, the X unit vector maintains the same direction, but the XZ plane is defined by the direction of the positive solar electric field, with the Z unit vector chosen to be along the electric field direction and orthogonal to the X and Y vectors. To define MSE coordinates, we use the Key Parameters (KP) version 12, revision 1 dataset, which includes data from the Particles and Fields instrument package onboard MAVEN \citep{Dunn2015}. The Solar Wind Ion Analyzer (SWIA) provides the proton flow velocity vector, while the magnetometer provides the IMF vector, giving us the background convection electric field from $\vec{E} = - \vec{v} \times \vec{B}$. However, due to its precessing orbit, MAVEN does not always sample the solar wind directly when its apoapsis is on the night side of the planet, so we remove those orbits from our study by examining the proton flow velocities and magnetic field signatures in the KP data, as has been done previously \citep{Halekas2017a}. Both proton flow velocities and magnetic field vectors are chosen to be the average values above 4000 km, to ensure that we are using solar wind values rather than those in the ionosphere.\\

To look at changes in average neutral density near the exobase for a variety of solar wind conditions, we compare densities for species O, Ar, and $\mathrm{CO_2}$ in the 180-220 km altitude range. This is generally at or just above the region where the atmosphere transitions from being dominated by collisions to becoming ballistic. By looking at this altitude range, we can examine neutrals that might be heated by incident particle flux to sufficient temperatures to affect the escape rate from the planet's gravity well. Ar was chosen because it is chemically inert, meaning it does not undergo photochemical processes, while O and $\mathrm{CO_2}$ were chosen due to their dominance at the altitudes of interest. For each orbit we average the measured densities for each species over altitudes 180-220 km. These averages are then separated into location bins of 5 degrees latitude and longitude in MSO and MSE coordinates, then the mean is found for each bin to produce the average density maps. This effectively normalizes the data by data density, reducing any bias due to multiple observations in the same location. Data is also split by solar longitude to examine seasonal neutral density variations outside of those expected from GCM models \citep{Bougher2014}. While we do not directly compute temperature, neutral heating can be inferred from higher average densities, as an increase in temperature will increase the scale height and hence we will see higher densities at a given altitude.\\

\section{Results} \label{results}
\begin{figure}[ht]
	\centering
	\includegraphics[width=25pc]{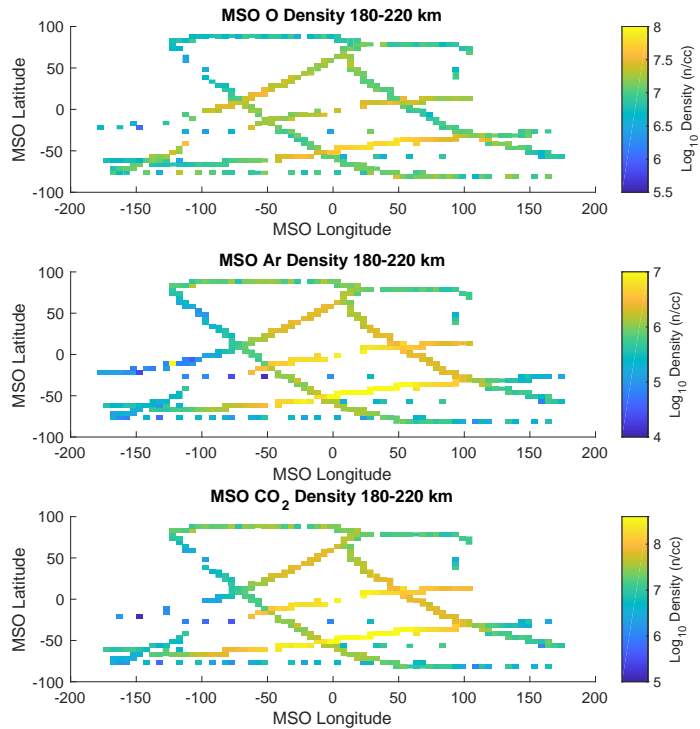}
	\caption{All available data averaged and binned in MSO coordinates with bin sizes of $5^{\circ}x5^{\circ}$ as described in the text: X pointing out from the plane towards the sun. Color indicates the log of the species density, with panels indicating density for $O$, $Ar$, and $CO_2$, respectively. Dayside, with noon at $0^{\circ}$, is uniformly higher density than the nightside (midnight at $180^{\circ}$), although the magnitude of the difference varies with species.}
	\label{MSOall}
\end{figure}

Figure \ref{MSOall} shows the binned densities as described in \ref{methods} in MSO coordinates. We observe, as expected, a distinct difference in density between dayside, the center of the figure, and nightside. For all three species, the density from subsolar point to antisolar point (or the data closest to those points) decreases by a significant amount. However, the highest density for O is not exactly in the noon region, as would be expected from atmospheric models in this altitude region \citep{Bougher2014}. For Ar and $\mathrm{CO_2}$, the difference between nightside and dayside at these altitudes is quite clear and is close to two orders of magnitude. This, not surprisingly, is consistent with a significant difference in average temperature due primarily to the effect of UV heating. The density gradient across the terminators ($\pm 90^{\circ}$) is relatively steep, dropping approximately an order of magnitude for argon and carbon dioxide across $50^{\circ}$ longitude. The density on the dayside and nightside is similar for the lighter species O due to ballistic transport and atmospheric winds. There are several places where the MSO paths cross, but do not have the same densities (for example, $0^{\circ}$ longitude, $-50^{\circ}$ latitude). This is due to changes in solar longitude between the passes.\\

\begin{figure}[ht]
	\centering
	\includegraphics[width=25pc]{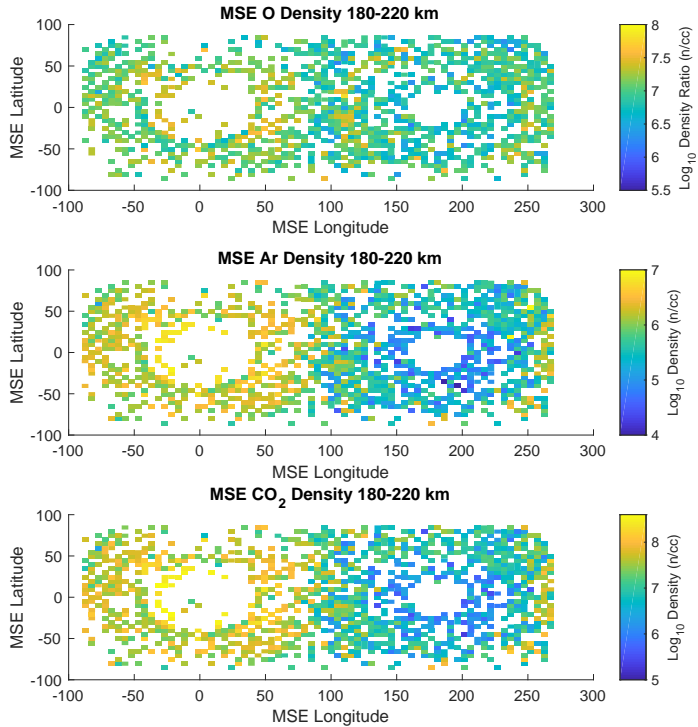}
	\caption{All available data binned and averaged with same resolution and color scale as in figure \ref{MSOall} but in MSE coordinates: X pointing out from the plane of each figure towards the sun and Z along projection of the solar wind electric field into the latitudinal plane, with Y perpendicular to the XZ plane and given as longitude. The solar wind electric field points away from the planet in the positive Z plane and towards the planet in the negative Z plane (positive and negative latitudes, respectively). Here, unlike in figure \ref{MSOall}, dayside is on the left side of the plot and nightside is on the right sides.}
	\label{MSEall}
\end{figure}

Figure \ref{MSEall} shows the same data as figure \ref{MSOall} but plotted in MSE coordinates. The MSE coordinates offer much higher spatial coverage than MSO, due to MAVEN retracing similar paths in MSO coordinates. While there are more filled bins, the number of orbits per bin on average is lower in figure \ref{MSEall}. The MSE coordinates for a particular region in MSO coordinates change frequently due to the transient nature of the solar wind and the flapping of the IMF. The highly variable Z coordinate in MSE gives much more latitudinal coverage resulting in the near azimuthal symmetry. Because of the higher coverage, the general trends observed in MSO coordinates for Ar and $\mathrm{CO_2}$ are easily seen: the dayside is higher density at this altitude than the nightside. However, any finer structure with respect to terminator or high latitude changes in density is lost because MSE coordinates are largely independent of geographic coordinates. The gap in the subsolar region is due to MAVEN not sampling the solar wind when its periapsis is at low solar zenith angles. At those times, its apoapsis is in the magnetotail or magnetosheath and so those data points are excluded, as the coordinate transformation will not be valid. The MSE coordinates give the densities an artificial oval or circular shape due to the rotation of the MSO coordinates with the solar wind electric field vector. So the circular spread represents rotation of similar locations in MSO for a variety of solar wind and IMF directions.\\

Previous papers have shown the presence of an ion polar plume at the north MSE pole \citep{Dong2015} as a significant source of ion escape. Because ions escaping from below the exobase can also heat this region of the atmosphere, such a feature would appear as a density enhancement in the northern MSE hemisphere. Additionally, ion precipitation into the atmosphere has been shown to be more common on the dayside hemisphere, specifically in the -E hemisphere \citep{Brain2015,Hara2017}, where the solar electric field points towards the planet. If this had a significant effect on the neutrals it would appear as an enhancement in the southern MSE hemisphere. Although the data density is not high, no such enhancements are obvious in figure \ref{MSEall}. Although it is possible that this would be more evident with additional data, these data suggest the effect of the incident plasma either does not, on average, significantly heat the neutrals in this altitude region or that the effect is uniformly distributed. The effect of the ions might be clearer if it was possible to have useful solar wind data when MAVEN's periapsis is on the dayside, thus providing data in the subsolar MSE coordinates region. Currently, the near noon dayside coverage is poorer than that on the nightside.\\

\begin{figure}[ht]
	\centering
	\includegraphics[width=25pc]{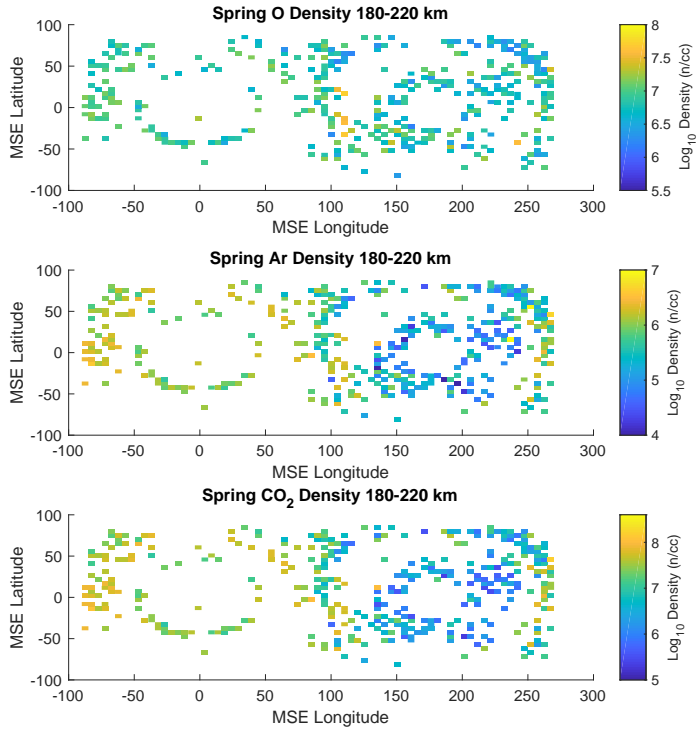}
	\caption{NGIMS data in MSE coordinates with a solar longitude less than $90^{\circ}$, i.e. northern spring, binned and averaged as described in the text. See figure \ref{MSEall} for a description of the figure.}
	\label{MSEspring}
\end{figure}

To pursue this further, in figure \ref{MSEspring} we bin the MSE density data for the Martian northern hemisphere spring and southern hemisphere fall, i.e. solar longitude less than $90^{\circ}$. Even though the data is even more sparse, for spring the coverage is best near the terminators and in the anti-solar region. We see the same general trends as in the entire dataset: dayside hemispheric densities are higher than the nightside densities for Ar and $\mathrm{CO_2}$. However, there is a suggestion that O seems to be higher density near $100^{\circ}$ longitude, which would be interesting as it matches model predictions quite well (e.g. \cite{Bougher2014}). This suggests that longitudinal trends in the density at this altitude are visible for the different species in MSE coordinates and that the enhancement is not an artifact of the data processing. The same local enhancement is roughly visible in Ar and $\mathrm{CO_2}$, which again would match global circulation models. However, any seasonal changes beyond those present in models for the neutral atmosphere near the exobase are not evident with the present data set.\\

\begin{figure}[ht]
	\centering
	\includegraphics[width=25pc]{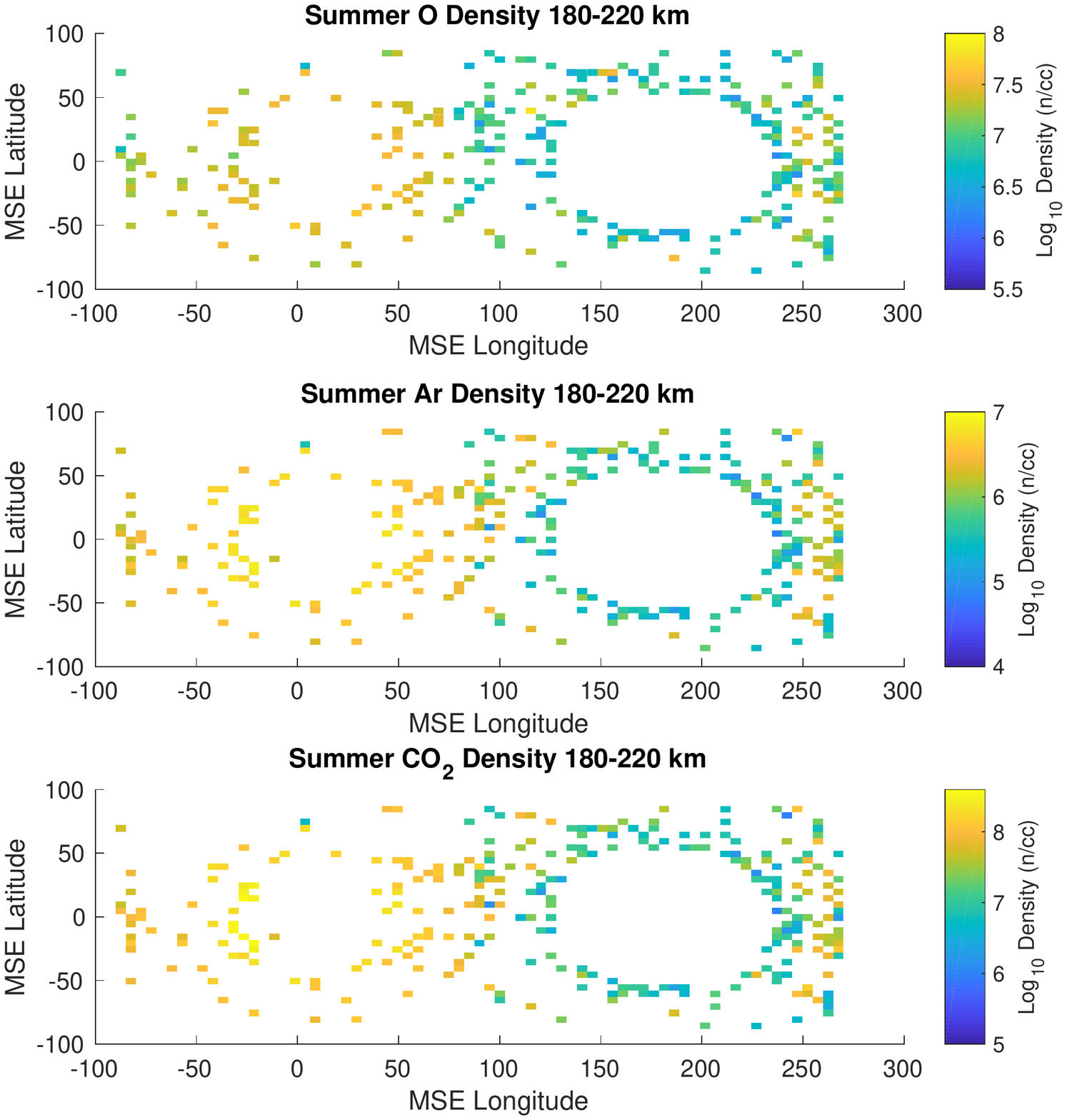}
	\caption{NGIMS data with a solar longitude between $90^{\circ} \leq L_s \leq 180^{\circ}$, i.e. northern summer, binned and averaged as described in the text. See figure \ref{MSEall} for a description of the figure.}
	\label{MSEsummer}
\end{figure}

Figure \ref{MSEsummer} shows the MSE binned data for northern summer and southern winter, which is also sparse. During this solar longitude, Mars is near its apoapsis. Again, overall longitudinal trends correspond roughly with those predicted by models. However, diurnal differences for O and $\mathrm{CO_2}$ are higher than predicted \citep{Bougher2014}. Specifically, the dayside for O is about an order of magnitude higher in density than the nightside in this altitude regime, contrary to what is seen in Mars global circulation models at similar altitudes. It is unclear why this might be the case; since this seems to be in both the +E and -E hemispheres, it is unlikely the precipitation is warming the dayside enough to increase the density by an order of magnitude. Therefore, it may be due to the coordinate transformation changing what the normal seasonal trends would look like. Unsurprisingly, the diurnal change in density is highest for this season. \\

\begin{figure}[ht]
	\centering
	\includegraphics[width=25pc]{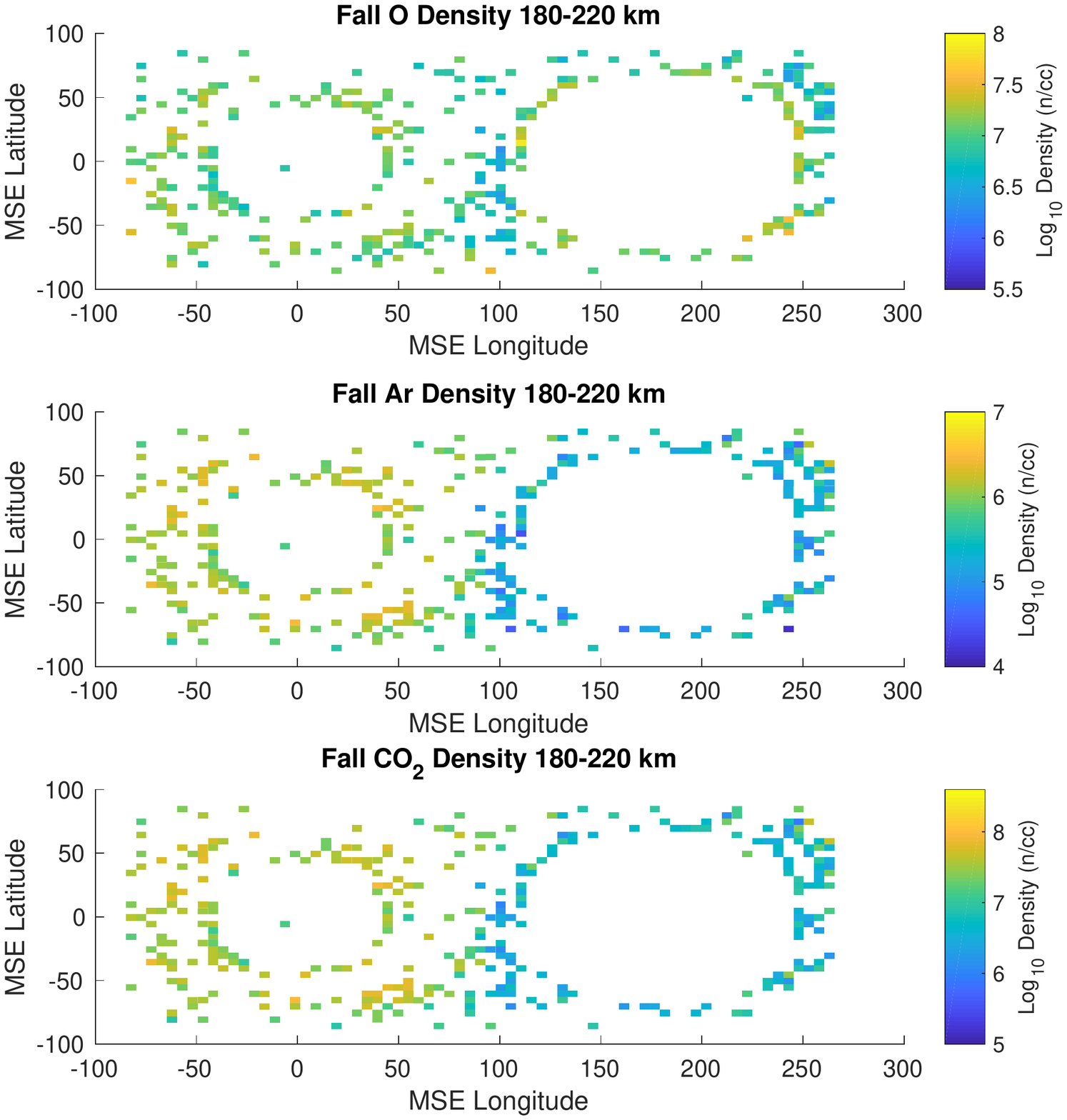}
	\caption{NGIMS data with a solar longitude between $180^{\circ} \leq L_s \leq 270^{\circ}$, i.e. northern fall, binned and averaged as described in the text. See figure \ref{MSEall} for a description of the figure.}
	\label{MSEfall}
\end{figure}

While data coverage is slightly different in figure \ref{MSEfall} for the fall equinox, the overall density trends are similar to figure \ref{MSEspring} at the spring equinox. The same high density region near $100^{\circ}$ and $250^{\circ}$ longitude for O is somewhat visible, although it is not as concentrated in a particular latitude region, indicating the variable solar wind conditions changing the MSE coordinates. $\mathrm{CO_2}$ and Ar densities are again as expected by GCM models. \\ 

\begin{figure}[ht]
	\centering
	\includegraphics[width=25pc]{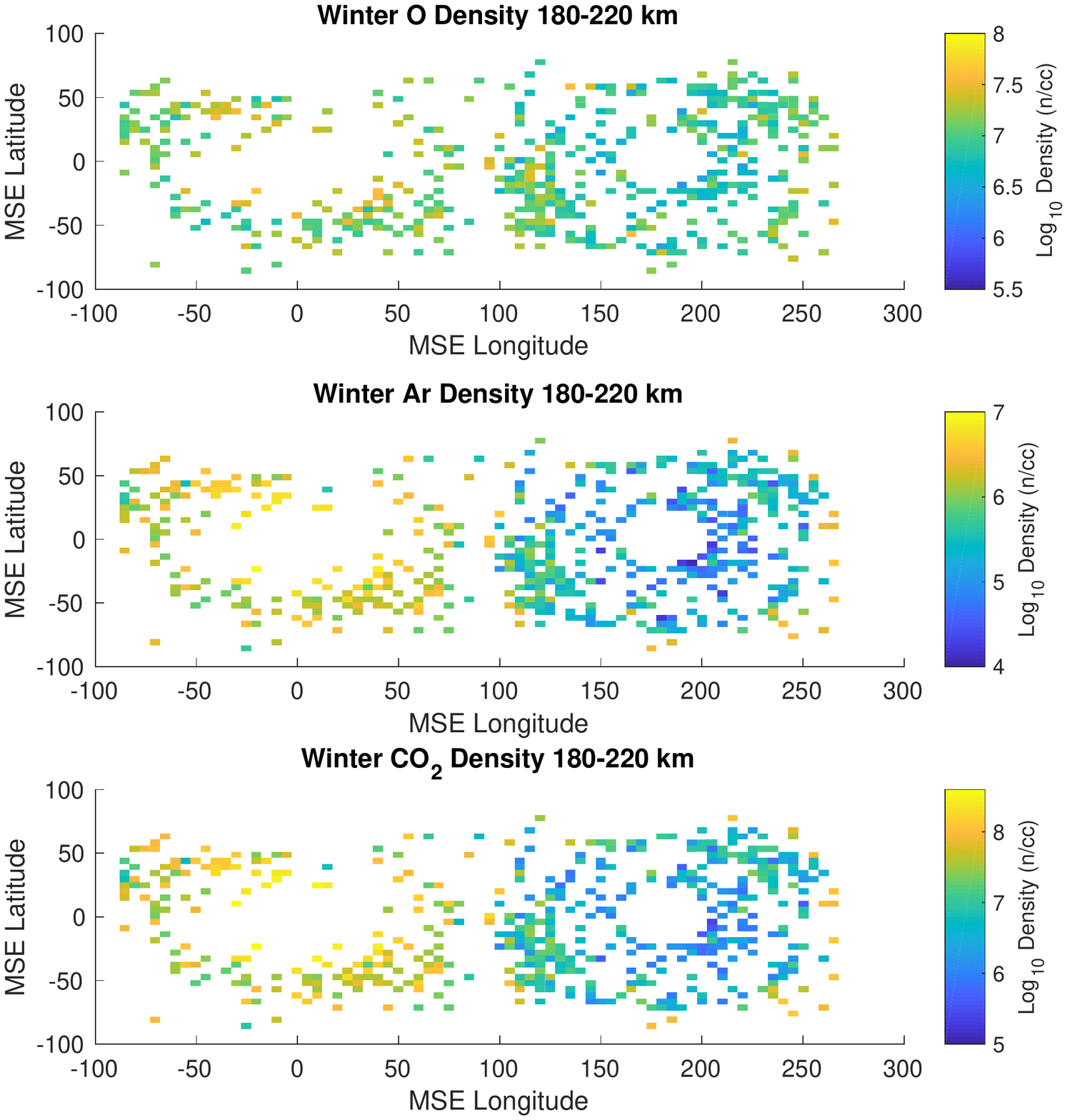}
	\caption{NGIMS data with a solar longitude greater than $270^{\circ}$, i.e. northern winter, binned and averaged as described in the text. See figure \ref{MSEall} for a description of the figure.}
	\label{MSEwinter}
\end{figure}

Because the start of the MAVEN mission was in Martian winter and the dataset used in this paper is slightly over a full Martian year, figure \ref{MSEwinter} is the only one of the four that contains data from more than one Martian year. Figure \ref{MSEwinter} shows data from northern winter, which is also the season closest to solar periapse and is typically dusty, which can warm or "puff up" the atmosphere. This season differs most from that predicted by GCM models, which predict that for O the nightside should be an order of magnitude higher density than the dayside at these altitudes, which we do not see here. Instead, both hemispheres are nearly equal in density. Likewise, Ar and $\mathrm{CO_2}$ are lower density on the nightside than would be expected. As mentioned above, the atmosphere close to periapsis can be quite dusty, which can change atmospheric temperatures and densities. GCM models typically use a dust average, usually for a weakly dusty season \citep{Bougher2014}. Thus is it likely that the difference between the data and model can be accounted for by dust, for while there has not been a global dust storm in some years, there are generally multiple large dust storms during this season every Martian year.\\

 \begin{table}
 \caption{Average densities for $\pm$E day and night hemispheres}
 \label{avgdens}
 \centering
 \begin{tabular}{| c | c | c | c |}
 \hline
 Hemisphere  & $O$ Density & $Ar$ Density & $CO_2$ Density\\
 \hline
   +E Day  & 1.2 & 0.14 & 4.5 \\
   -E Day  & 1.4 & 0.15 & 4.6 \\
   +E Night  & 0.94 & 0.07 & 1.9 \\
   -E Night  & 1.1 & 0.08 & 2.4 \\
 \hline
 \multicolumn{2}{l}{$^{a}$*1.0e7/$cm^3$}

 \end{tabular}
 \end{table}

In table \ref{avgdens}, we have split the MSE density data into dayside positive and negative E field direction (i.e. latitude) and nightside positive and negative E field direction, then taken the mean for each group. This shows that for all three species, regardless of solar zenith angle, the -E hemisphere is slightly higher in density than the +E hemisphere. Therefore it is possible that ion precipitation is adding energy to the neutral atmosphere in the -E hemisphere, hence raising the temperature and expanding the atmosphere. However, the standard deviation for these averages is large, up to 50 percent of the average density value, due to the data being from an entire hemisphere and thus varying widely in density. As a result, the differences between the $\pm$E hemispheres are not individually statistically significant, but are consistent for all three species, as is the magnitude of the change, suggesting that precipitation might be affecting the neutrals on a global scale.\\

\section{Discussion} \label{disc}

Both diurnal and seasonal local time variations in density near the exobase are easily visible in the MSE coordinates and show the expected changes. Seasonal variations are more complex, since they also depend on geographic latitude, not just solar longitude. However, the data shown in MSE coordinates is consistent with those studies in which the data is examined in geodetic coordinates. The orbital distance of Mars from the Sun, of course, also plays a role as it is highly eccentric affecting the solar insolation and, hence, the atmospheric structure. Comparing figure \ref{MSEsummer} and figure \ref{MSEwinter} shows densities at solar apoapse are lower than those near periapse as expected \citep{Bougher2014,Bougher2015}.\\

By definition, MSE coordinates provide information about the effect of ion flow in various regions. Because models have demonstrated that ion precipitation can transfer energy and increase neutral temperatures \citep{Fang2013,Michael2005}, which in turn causes an increased scale height, comparing densities at the same altitude could be a proxy for the upper atmosphere heating rate. This is reflected the seasonal effect and higher average dayside densities at fixed altitude seen in figures \ref{MSOall} and \ref{MSEall} for Ar and $\mathrm{CO_2}$ \citep{Valeille2009,Mahaffy2015,Lillis2015}. Therefore, the slightly higher densities seen in the -E hemisphere in table \ref{avgdens} suggests it is on average, slightly warmer than the +E hemisphere for both day and night. \\

Consistent with other studies using MAVEN data, atmospheric heating by incident plasma ions is found to be a small effect during the period examined. Our analysis suggests that heating of the neutrals on entering (-E hemisphere) slightly dominates heating on exiting (+E hemisphere). Since the change in density from MSE north to south in the altitude region studied is within one standard deviation of the mean, it is not statistically significant. However, the observed few percent difference is consistent with recent models of ion precipitation and sputtering \citep{Fang2013,Leblanc2017}. Due to the expected size of the density change and the lack of coverage on the dayside, it is not obvious in the global density maps (e.g. figure \ref{MSEall}). To further evaluate its importance will require data at times of high solar activity, as well as data at lower solar zenith angles, which is not possible without independent solar wind measurements made simultaneously with the density measurements. Although there is a suggestion of an effect, the role of pick-up ion precipitation, thought to be critical in the evolution of the Martian atmosphere, remains elusive.

%
%
%
%
%
%
%

\acknowledgments
Support at the University of Virginia comes from NASA's Planetary Data Systems Program via grant NNX15AN38G. The data used in this study is available from the NASA Planetary Data Systems Atmospheres Node and the MAVEN Science Data Center at the University of Colorado Boulder Laboratory for Atmospheric and Space Physics.

\listofchanges

\end{document}